\documentclass[twocolumn,twoside,slac_two]{revtex4}

\usepackage{graphicx}
\usepackage{fancyhdr}
\def\d0    {D\O}								

\def\met   {$\not E_ \bot $}				
\def\lum   {$\cal{L}$}						
\def\lumunits{cm$^{-2}$s$^{-1}$}			
\def\met   {\mbox{${\hbox{$E$\kern-0.6em\lower-.1ex\hbox{/}}}_T$ }}	
\begin{document}

\title{Results from the Commissioning of the ATLAS Pixel Detector with Cosmic data.}
\author{E. Galyaev}
\affiliation{Department of Physics, University of Texas at Dallas, Richardson, TX 75080, USA}
\begin {abstract}
The ATLAS pixel detector is the innermost detector of the ATLAS experiment at the Large Hadron Collider at CERN. With approximately 80 million readout channels, the ATLAS silicon pixel detector is a high-acceptance, high-resolution, low-noise tracking device. Providing the desired refinement in charged track pattern recognition capability in order to meet the stringent track reconstruction requirements, the pixel detector largely defines the ability of ATLAS to effectively resolve primary and secondary vertices and perform efficient flavor tagging essential for discovery of new physics.
 
Being the last sub-system installed in ATLAS by July 2007, the pixel detector was successfully connected, commissioned, and tested {\it in situ} while meeting an extremely tight schedule, and was ready to take data upon the projected turn-on of the LHC. Since fall 2008, the pixel detector has been included in the combined ATLAS detector operation, collecting cosmic muon data. Details from the pixel detector installation and commissioning, as well as details on calibration procedures and the results obtained with collected cosmic data, are presented along with a summary of the detector status.
\end{abstract}
\maketitle
\thispagestyle{fancy}
\section{Introduction}
ATLAS ({\it A Toroidal Large Acceptance Spectrometer}) is one of two multi-purpose particle detector experiments designed to explore physics produced at the energy frontier with colliding proton beams at the LHC. Built around one of the crossing points for LHC beams, ATLAS utilizes a highly granular, multi-layered detector construction, based on nearly hermetic $4\pi$ geometry with respect to the collision region. The inner tracking volume, immersed in a 2~T solenoidal magnetic field, begins just a few centimeters away from the proton beam axis, extends to a radius of 1.1 meters, and is seven meters in length along the beam pipe. Its basic function is to track charged particles by detecting their interactions with the detecting media at discrete points with high precision, revealing detailed information about the type of particle and its momentum. The inner tracker is comprised of three sub-detector layers, nested from the interaction point to the periphery: a silicon pixel detector, a semi-conductor tracker, and a transition radiation tracker.

The operational conditions impose stringent requirements on the overall pixel detector design. The expected dose for the innermost layer is assumed to reach 500 kGy within approximately five years of LHC operation. Minimum amounts of materials are used for all elements of the pixel detector in order to reduce multiple scattering and secondary interactions.

Being the innermost tracking sub-system, the pixel detector provides excellent impact parameter resolution and low occupancy per readout channel. Its design defines the ability of ATLAS to identify and reconstruct secondary vertices from the decay of long-lived particles containing heavy quarks, or for flavor-tagging of jets with extremely high overall track multiplicity per event. In addition, the pixel detector provides excellent spatial resolution for reconstructing primary vertices close to the proton-proton interaction region within the detector, with multiple interactions being common at full LHC design luminosity of \lum $ =10^{34}$~\lumunits.
\section{Overview of the Pixel Detector}
\begin{figure}[h]
\centering
\includegraphics[width=80mm]{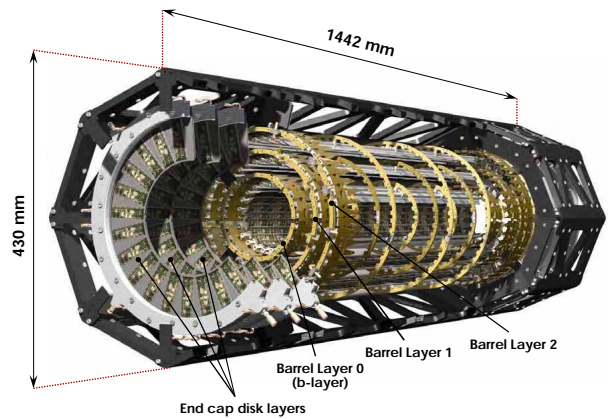}
\caption{General view of the ATLAS pixel detector.} 
\label{fig:01_Pixel_Overview}
\end{figure}
The ATLAS silicon pixel detector (Fig.\ref{fig:01_Pixel_Overview}) contains approximately 80 million readout channels in total. The mechanical detector layout consists of three concentric cylindrical layers located respectively at radii of $50.5~\mathrm{mm}$, $88.5~\mathrm{mm}$ and $122.5~\mathrm{mm}$ from the beam axis, and two symmetrical end caps with three disks in each, attached to both sides of the barrel region. Such layout ensures the design requirement of having at least three pixel hits in the pseudorapidity range of $|\eta|<2.5$. The 1744 identical pixel modules, each measuring roughly as two centimeters by six centimeters, provide a total sensitive area of $\sim 1.7~\mathrm{m^2}$. The modules are located on lightweight carbon support structures, with integrated $\mathrm{C_{3}F_{8}}$ active bi-phase evaporative cooling in place, in order to keep modules at the desired operational temperature of about $0^{\circ}$C to minimize irradiation effects and noise. The identical mechanical supports in the barrel layers are called staves. The three barrel layers are built respectively with 22, 38, and 52 staves, each holding 13 pixel modules. The disks are made up of sector structures, with each sector holding six modules. Each complete disk is comprised of eight sectors. Fully assembled, the pixel detector is sized $34.5~\mathrm{cm}$ in diameter by 1.3~m in longitude, weighting only 4.5~kg. A more detailed description of the ATLAS pixel detector can be found in~\cite{bib:01_pixel_tdr},\cite{bib:02_electronics_sensors}.
\subsection{The Pixel Module}
\begin{figure}[h]
\centering
\includegraphics[width=80mm]{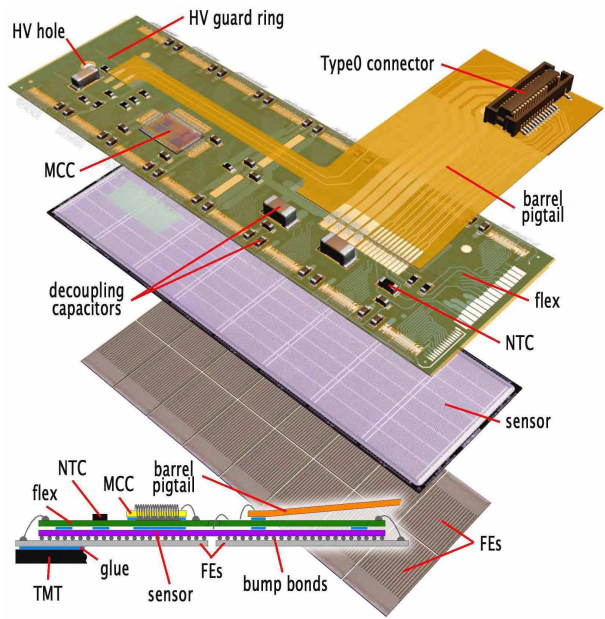}
\caption{Schematic view of the construction and components of a pixel module~\cite{bib:02_electronics_sensors}.} 
\label{fig:02_Pixel_Module}
\end{figure}
\begin{figure}[ht]
\centering
\includegraphics[width=80mm]{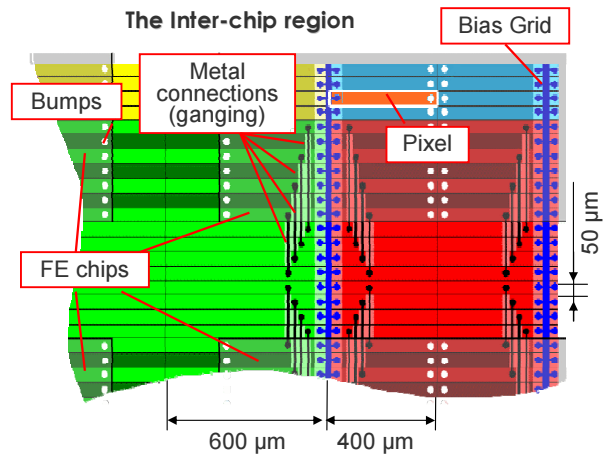}
\caption{Schematic view of the inter-chip region in a module~\cite{bib:02_electronics_sensors}. Pixels in regions between the FE chips are being linked in pairs with the ones covered by active readout electronics, or {\it ganged}.} 
\label{fig:03_Ganging}
\end{figure}

The basic building block of the active part of the pixel detector is a pixel module (Fig.\ref{fig:02_Pixel_Module}) that is composed of silicon sensors, front-end electronics, and flex-hybrids with integrated control circuits~\cite{bib:02_electronics_sensors}. All modules are functionally identical at the sensor and at the integrated circuit levels, but differ somewhat in design of the connection points for barrel modules and for disk modules. In order to achieve the nominal design specifications on impact parameter resolution of $15~\mathrm{\mu m}$ in the $r-\phi$ direction, the nominal pixel size is 50 microns in the $\phi$ direction and 400 microns in $z$ (barrel region, along the beam axis) or $r$ (disk region). There are 46,080 pixel electronics channels in each module, being read out by 16 front-end ({\it FE}) chips, arranged in 2 columns of 8 chips, which are attached to the sensor pixels by either In or PbSn bump bonds. Approximately 13\% of the 47232 physical pixels on each sensor have slightly larger dimensions of $50\times 600~\mathrm{\mu m}$ to allow for a contiguous sensitive area between the FE chip boundaries in the longitudinal pixel direction. In the transverse direction, $2\times 4$ pixels under each of the two adjacent chips cannot be covered by active pixel circuitry. These special pixels are interconnected, or {\it ganged} by the conductive traces on the sensor, with one of $4+4$ neighboring electronics pixels on top of the columns, as is illustrated in Fig.\ref{fig:03_Ganging}. The resulting hit ambiguity is resolved by the off-line pattern recognition algorithm~\cite{bib:02_electronics_sensors}.

Sensors are the sensitive part of the pixel detector used for charged particle detection, and their function can be loosely compared to microscopic solid-state ionization chambers. The ATLAS pixel sensors are arrays of bipolar diodes placed on a high resistivity, $250~\mathrm{\mu m}$ thick $n$-type bulk, with $n+$ implants on the readout side, and a $p-n$ junctions on the other side of the sensor. The oxygen-enhanced silicon, so-called {\it Diffusion-Oxygenated Float Zone} ({\it DOFZ}), is used, as it has been proven to provide more tolerance to irradiation than the standard float-zone silicon~\cite{bib:02_electronics_sensors}. The sensors are operated in reverse bias, with full depletion voltages ranging from $\mathrm{150~V}$ to $\mathrm{600~V}$, depending on the $p-n$ junction conditions, determined by the irradiation damage.

Each of the FE readout chips of the ATLAS pixel detector contains 2880 individual readout cells. There is an analog and a digital circuitry in every pixel cell. The charge created in a pixel by a traversing charged particle is collected and amplified by in-cell charge-sensitive preamplifier in the analog part of a cell's electronics. To reduce the data volume transferred out of the detector, only those hits that produce a charge above an individually adjustable predefined threshold, are read out. The individual threshold level is set in a discriminator at the output of each cell's preamplifier. The hits above threshold are digitized before they are collected in a memory buffer at the front-end chip periphery, including the hit address, the signal's leading and trailing edge time-stamps, and the collected charge measurement determined by the {\it Time-Over-Threshold} ({\it ToT})\footnote{ToT is measured in units of {\it Bunch~Crossings} ({\it BC}s). \newline \indent \indent BC equals to 25~ns.}, and stored there for up to $\mathrm{6.4~\mu s}$. Pixel cells are being read-out in columns, with each double-column operated by a separate {\it Controller End-of-Column} ({\it CEU}) unit. The readout buffers with 64-hit capacity are available within each CEU. If the time-stamp of the hit matches with a Level-1 trigger signal, the hit is flagged for readout and sent to the {\it Module Controller Chip} ({\it MCC}), located on a flex-hybrid printed circuit board, glued to the back of the sensor. The MCC manages {\it Timing, Trigger and Control} ({\it TTC}) signals to the front end chips. The MCC can replicate a trigger signal up to 15 times, which allows for readout windows of up to $\mathrm{400~ns}$. The flex-hybrid circuit of the pixel sensor also houses the decoupling capacitors and the {\it Negative Thermal Coefficient} ({\it NTC}) sensor for temperature measurement. The barrel modules possess an additional flexible layer - a {\it pigtail}, with a connector for a low-mass cable. This low-mass cable connects modules to the {\it Patch Panel~\O} ({\it PP\O}), while the disk modules are connected without using the pigtails~\cite{bib:02_electronics_sensors}.
\subsection{Pixel Detector Readout}
\begin{figure}[h]
\centering
\includegraphics[width=80mm]{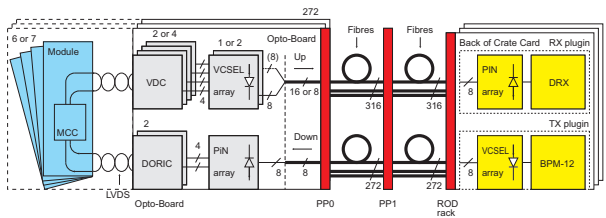}
\caption{Diagram of the communication links between pixel modules and off-detector readout electronics~\cite{bib:02_electronics_sensors}.} 
\label{fig:04_Optolinks}
\end{figure}

In the ATLAS pixel detector, optical communication links are used to pass on data and control signals between modules and off-detector readout electronics. The overall schematics of the optical readout communication links is presented in Fig.\ref{fig:04_Optolinks}.
The optical links are custom made, and are implemented using commercial-grade, expitaxial silicon-based PIN diodes for the optical receivers as well as {\it Vertical Cavity Surface Emitting Laser} ({\it VCSEL}) arrays bare die, with custom-designed integrated readout control circuits. On-detector optical communication is realized by means of the optical link boards (the $opto-boards$), and via the {\it Back of Crate} ({\it BOC}) communication cards on the off-detector end.

The transmission of the signals from the detector modules to the opto-boards uses {\it Low Voltage Differential Signaling} ({\it LVDS}) electrical connections. These serial connections link the MCC with the {\it VCSEL Driver Chips} ({\it VDC}s) and the {\it Digital Optical Receiver Integrated Circuits} ({\it DORIC}s), instrumented on the opto-board PCBs made with beryllium oxide. The communication with each detector module uses individual fibers: one for the downlink and one (or two for the B-Layer boards) for the uplinks. About $\mathrm{80~m}$ of radiation-resilient clear optical fibers interconnect 132 BOC cards located in racks inside the ATLAS cavern, and 272 opto-boards on the detector side. The number of connections per BOC depends on the readout speed, which at the full LHC luminosity is designed to be $\mathrm{160~Mbit/s}$ for the B-Layer, $\mathrm{80~Mbit/s}$ for the Layer-1 and the disks, and $\mathrm{40~Mbit/s}$ for the Layer-2 modules. Trigger, clock, commands and configuration data travel on the downlink, while event data and configuration read-back data travel on the uplink(s). On the downlink, a {\it Bi-Phase Mark} ({\it BPM}) encoding is used to send a $\mathrm{40~Mbit/s}$ control stream on the same channel as the $\mathrm{40~MHz}$ bunch crossing clock. Decoding of the BPM channel within the DORIC recovers both the data stream and the clock signal. The use of individual links for every module permits the adjustment of the timing used to associate the hit to a bunch crossing. The timing adjustment is accomplished by changing the delay of the transmitted signal with respect to the phase of the LHC machine reference clock received in the BOC~\cite{bib:02_electronics_sensors}.

The differential data output of the module is converted into single-ended current signals in the VDC, driving each VCSEL array. Output power of the laser arrays is controlled by the $V_{I_{set}}$ voltage, which has the same value for all channels on each opto-board. Data are received by the Rx plug-in of the BOC on the off-detector side, where the phase between the data and the sampling clock are being adjusted for every communications channel~\cite{bib:03_electronics_calibration}. It was found that the opto-boards require to be maintained at a certain operating temperature for stable, error-free operation. A special system of opto-heaters was designed and integrated on-detector for that purpose.

The pixel detector data are processed by 132 {\it ReadOut Drivers} ({\it ROD}s), each being a counterpart to the BOCs in the readout crates.~Each of the total of nine readout crates additionally houses one {\it Single Board Computer} ({\it SBC}) and one {\it TTC Interface Module} ({\it TIM}) that interfaces a TTC crate. This crate, organized into three partitions (corresponding to the B-Layer, the Layer 1 and the Layer 2 combined, and the Disks) distributes the TTC signals from the {\it Central Trigger Processor} ({\it CTP}) and collects the BUSY signal from the RODs, which, being transmitted to the CTP, can stop global ATLAS data acquisition. The core of each partition is a {\it Local Trigger Processor} ({\it LTP}), which can also generate the TTC signals locally.~This capability is used in standalone data-taking runs.~The readout of the pixel detector can be operated in the data-taking mode or in the calibration mode.~In the data-taking mode, the RODs receive Level-1 triggers from the TTC crate and build local events from the received raw pixel hits data. These events are sent through the S-Link to the {\it ReadOut System} ({\it ROS}), where, depending on the Level-2 trigger decision, they will be either included into the ATLAS event or discarded~\cite{bib:02_electronics_sensors},\cite{bib:05_iskander_proceedings}.

In the calibration mode, trigger and control signals are generated in the RODs, and module events are sent through the VME bus to the SBC for further processing. Since the amount of data generated in calibration procedures is ample, {\it Digital Signal Processors} ({\it DSP}s) are employed on the RODs~\cite{bib:02_electronics_sensors}. The typical DSP processing tasks are histogramming, averaging and fitting the acquired module data, thus decreasing considerably the time that the calibration procedures require~\cite{bib:03_electronics_calibration}.
\section{Commissioning and Calibration}
The assembly of the pixel detector package was completed in spring 2007, and the package was inserted into the center of the ATLAS inner detector at the end of June 2007~\cite{bib:04_status_overview}.~The package included the completed pixel detector along with its service connections, mounted around the central section of the ATLAS beryllium beam pipe. Shortly thereafter, physical access to the pixel detector assembly became extremely limited due to other commissioning tasks on the inner detector. During the following six months, all of the supplementary electrical services were routed and installed.~A series of tests were performed to ensure proper functionality of the opto-heater control system. All the necessary electronic components for opto-board temperature control have been assembled and tuned using custom-built dummy loads. In winter of 2007/08, the pixel opto-heater system was tested in an environment closely replicating the combined pixel operation conditions by means of using the test setup located in the experimental area nearby the ATLAS cavern.~At the same time, in fall 2007, the off-detector readout electronics were installed in the counting rooms in the cavern.
\subsection{Initial Connection of Services and Cooling}
\begin{figure}[htp]
\centering
\includegraphics[width=80mm]{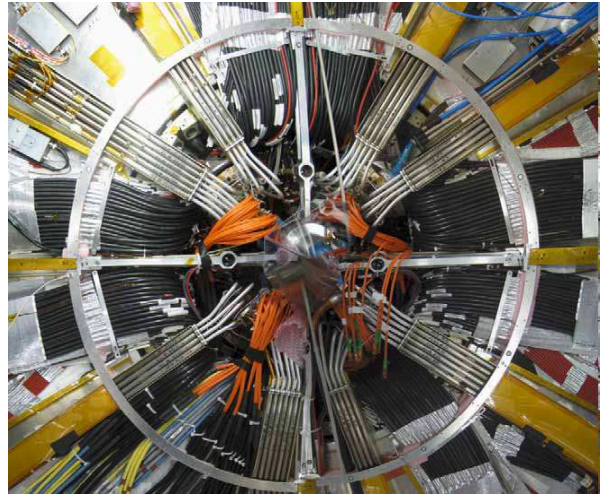}
\caption{A photograph of the PP1 area with the pixel detector connected. Photo courtesy of ATLAS Pixel Collaboration.} 
\label{fig:05_PP1}
\end{figure}

Initial connection of the pixel detector package to the cooling and electrical services, and to the off-detector readout electronics began in February 2008, after finalizing the connections of the other ATLAS sub-detectors. Full connection, accompanied by a series of connectivity and performance checks, was accomplished by April 2008.~A photograph of the {\it Patch Panel 1} ({\it PP1}) area (only 0.5m in diameter) after the connections were made is shown in Fig.\ref{fig:05_PP1}.
During the sign-off period, the cooling loops were commissioned by a variation of the heat load, using special module configurations. Some loops showed instabilities when the detector was switched off; however, after several modifications of the cooling regulation, the cooling system has been working more reliably. In the course of spring 2008, the opto-board heater system has been deployed, connected, and tested, along with the cooling system.

A cooling plant failure happened on May 1, 2008, which prevented the timely sign-off of the detector. Three out of six compressors that generate pressure to liquefy the cooling gas did not startup correctly after a sudden stop of the plant.~The magnetic mechanical couplers between the motor and the pump kept slipping for several hours, causing subsequent overheating and failure.~As cracks developed in the seals of the magnetic couplers, contamination of the $\mathrm{C_{3}F_{8}}$ coolant with pump oil occurred, however no substantial contamination was found in the pixel detector's structures. To lower the risk of similar failures, after the incident the compressors were equipped with auxiliary sensors and in-line coolant filters~\cite{bib:06_sara_proceedings}.

Commissioning of the cooling plant was completed by the end of August 2008, with all 88 cooling loops being operable. Three cooling loops in the pixel detector end caps were found to be leaky and, after re-calibrating the respective modules, were switched off until the winter 2008/09 shutdown. Further leak rate measurements were made during the shutdown. The problematic cooling loops have been repaired and were put back in operation in May 2009.
\subsection{Tuning of the Optical Connections}

\begin{figure}[htp]
\centering
\includegraphics[width=85mm]{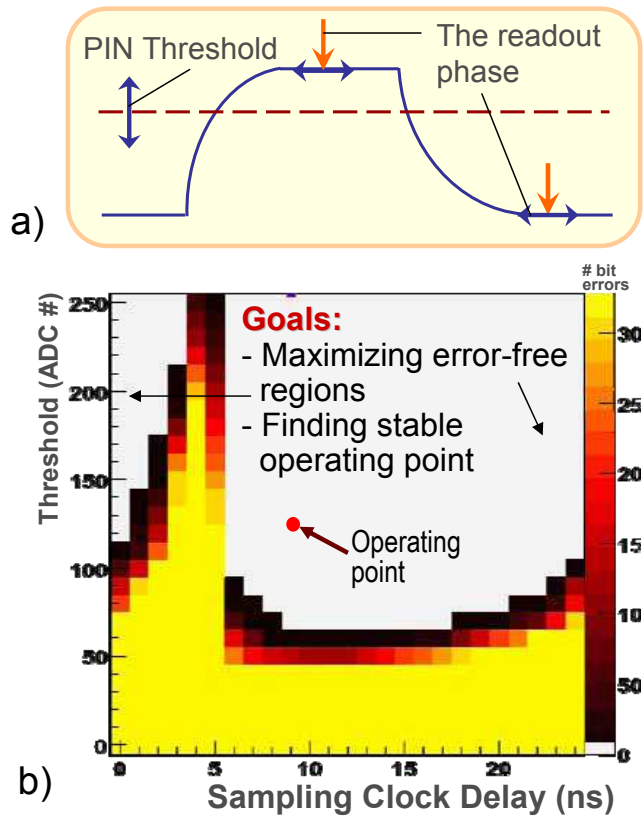}
\caption{a) Illustration of the parameter space used to tune the opto-links. b) Scan of the parameter space of the Rx plug-in in the course of the optical tuning. White areas of the plot represent error-free regions, while the color scale reflects the number of bit errors.} 
\label{fig:06_BOCscan}
\end{figure}

To ensure stable operation of the pixel detector, reliable and error-free optical communication has to be established with all of the pixel modules. The optical communication tuning procedure has three parts: confirmation of the downlink, verification of the uplink, and tuning of the uplink~\cite{bib:03_electronics_calibration}. Confirmation of the downlink is accomplished by sending light from the Tx plug-in to the off-detector BOC card through the downlink fiber, and measuring the current of the PIN diode on the opto-board. Verification of the uplink is done by issuing a command to the modules to send a $\mathrm{20~MHz}$ clock test pattern through the uplink fibers to the BOC. At the BOC, the signal is received and measured by the PIN diode in the Rx plug-in.~Tuning of the optical uplink includes adjusting threshold and delay of the Rx plug-in, as well as varying the output power of the VCSELs on the opto-boards.~The pixel modules are called to send out a $\mathrm{20~MHz}$ clock pattern to the BOC. The threshold and delay values at the BOC are adjusted until the clock pattern is correctly recognized (Fig.\ref{fig:06_BOCscan}).

The power of the emitting VCSEL is required to stay within the limits set by the off-detector threshold range, at the same time providing a stable signal. Certain channels, especially those in which the VCSELs are exhibiting {\it slow turn-on behavior}\footnote{VCSEL on the opto-board require $10\sim \mathrm{250~\mu s}$ to reach full output power.}, require a repeated pass of tuning. In this second pass, the modules are required to send back a pseudo-random test data pattern, which is received by the Rx plug-in inside the BOC. New values for threshold and delay are chosen within the error-free region for these problematic channels.~The on-detector VCSEL power can be readjusted as well to obtain optimal link configuration.~To confirm the success of the optical tuning procedure and to verify the functionality of the digital part of the FE chip, a digital pulse is injected into each readout channel after the discriminator via the integrated in-cell charge generator.~If such simulated hit is correctly registered by the BOC, it concludes the optical tuning procedure, as well as ensures the correct digital response of the FE chip.\begin{figure*}[ht]
\centering
\includegraphics[width=135mm]{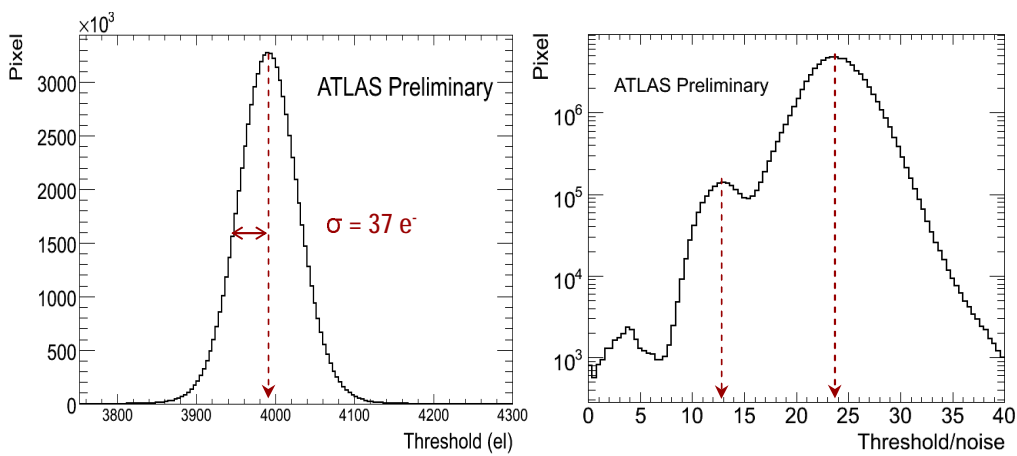}
\caption{a) The distribution of the tuned thresholds for the majority of pixels in the detector. b) Distribution of the threshold-to-noise ratio in most pixels.} \label{fig:07_Thresholds}
\end{figure*}

As it has been established in the course of tests performed in 2006, amongst the critical parameters in reliable operation of the optical readout links is the temperature of the opto-board~\cite{bib:07_system_test}. Thus, the additional environmental control system was introduced to keep the favorable temperature conditions of the opto-boards during detector operation in the actively cooled pixel detector surroundings. The total of 48 miniature resistive heater strips, paired with the NTC sensors, were attached to the opto-boards to keep their temperature constant in the course of operation and ensure the stability of the communication link parameters.

It is desirable to keep the temperature of the opto-boards and adjacent components as low as possible in order to reduce adverse irradiation effects. Therefore as soon as the active cooling became available, optical links were tuned at $10^{\circ}$C. Since several channels have exhibited a slow turn-on behavior, it was decided to increase the operating temperature of all opto-boards to $20^{\circ}$C~\cite{bib:05_iskander_proceedings}. At the moment, about 97\% of the optical connections are being tuned using the above calibration procedure, while a few remaining links require individual approach~\cite{bib:03_electronics_calibration}.  
\subsection{Analogue Calibration and Performance}
The analogue performance of a pixel is measured using an injection circuit, which injects a series of pulses of known charge into the preamplifier via in-cell integrated circuitry.~By repeating the injection and scanning the range of applied charge, one can measure the discriminator threshold and the noise level of the channel.~This so-called {\it threshold scan}, is also employed for validation of the threshold tuning~\cite{bib:03_electronics_calibration}. The current tuning algorithm injects a charge exactly at the aimed threshold and then alters the settings of the readout channel in the front-end chip until the occupancy of that pixel is 50\%. While the most probable charge deposition in a pixel from a {\it Minimally Ionizing Particle} ({\it MIP}) is $\sim \mathrm{20~Ke^-}$, the thresholds are being tuned pixel by pixel to the target value of $\mathrm{4~Ke^-}$. The resulting turn-on curve is convoluted with the noise in the pixel, which is approximated by a Gaussian distribution, and then the resulting dependence is fitted to an error function. The mean of the fitted curve corresponds to the threshold value of $\mathrm{3,95~Ke^-}$, and the curve width reflects the noise. With the present threshold-tuning algorithm, it is possible to tune over 95\% of the detector, having a threshold dispersion of just $\mathrm{37~e^-}$ (Fig.\ref{fig:07_Thresholds}).
 
The threshold-to-noise ratio was measured to be close to 24 for most of the pixels for the target threshold of $\mathrm{4~Ke^-}$, whereas for the few pixels in the inter-chip region it was approximately 13, with twice as much noise observed in them, resulting from having higher input capacitance due to being interconnected, or {\it ganged}~\cite{bib:02_electronics_sensors},\cite{bib:06_sara_proceedings}.

After having balanced the noise threshold setting of the discriminator, another part of tuning procedure adjusts the preamplifier gain in order to respond uniformly to a certain input charge by regulating the preamplifier feedback current. The preamplifier uses a programmable current in the feedback loop so that the ToT is a nearly linear function of the deposited charge. For each hit above threshold, the time-stamp of when the leading and falling edges of the signal cross the threshold are registered, in units of bunch crossings. The difference between the time-stamps of the falling and the leading edges is defined as time over threshold. The leading edge time stamp defines the bunch crossing a hit is registered in. By injecting the $\mathrm{20~Ke^-}$ MIP-like charge into the preamplifier, the pixel response is tuned to correspond to ToT of 30 BCs~\cite{bib:02_electronics_sensors},\cite{bib:03_electronics_calibration},\cite{bib:06_sara_proceedings}.
\subsection{Noise Occupancy Measurement}\begin{figure}[!htbp]
\centering
\includegraphics[width=85mm]{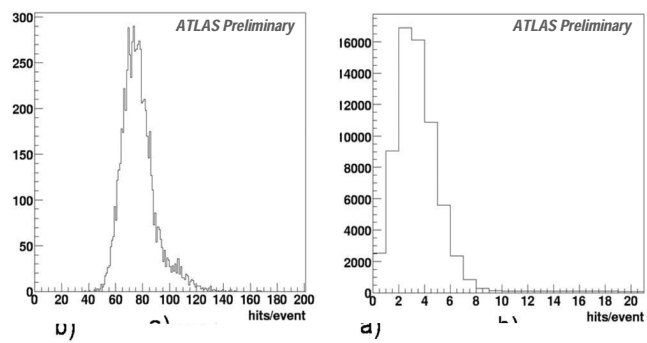}
\caption{Number of hits per event before (a) and after (b) masking of noisy pixels~\cite{bib:05_iskander_proceedings}.} 
\label{fig:08_Masking}
\end{figure}
\begin{figure*}[!htbp]
\centering
\includegraphics[width=135mm]{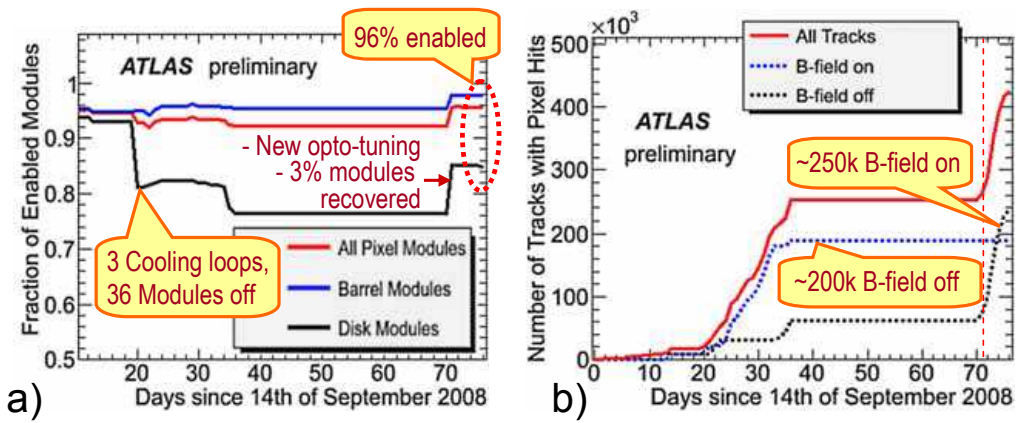}
\caption{a) Fraction of modules included in cosmic runs in fall 2008. Numbers for different parts of the detector (and containing different overall number of pixel modules) are normalized to 100\% separately. b) The number of tracks with at least one pixel hit as a function of time.} 
\label{fig:09_Running}
\end{figure*}
Noise occupancy is measured by reading out the pixel detector randomly, with no reconstructed tracks observed.~For this purpose, stand-alone data taking runs with a trigger frequency of a few kHz are used. After stable data taking is achieved and data samples recorded, noise occupancy per pixel is calculated and the noisy pixels map, containing only the pixels, which had occupancy greater than $10^{-5}$, is created.~Presently, the map contains less than 5000 of these ``hot" pixels, which are being masked. After the masking, the noise occupancy dropped by a factor of 20, as shown in Fig.\ref{fig:08_Masking}, to about 0.5 hits per event per beam crossing for the entire pixel detector~\cite{bib:06_sara_proceedings}.~This corresponds to about $2\times 10^{-10}$ noise occupancy per pixel per BC for the whole detector, or less than one noise hit per event in $\mathrm{80~M}$ channels. The reduction of noise by applying the noise mask, and the stability of the results over time confirms the fact that most of the noise comes from the same pixels. The noise mask is being updated on a regular basis, and after masking only 0.006\% of the pixels, the noise level is very low.
\section{Cosmic Ray Data Taking}
The pixel detector was included in combined data taking with the rest of ATLAS for the first time on September $\mathrm{14^{th}, 2008}$~\cite{bib:04_status_overview}. Upon the initial adjustment of trigger timing, first cosmic muon tracks through the pixel volume were recorded.~When single beam was first injected into the LHC on September $\mathrm{10^{th}}, 2008$, the pixel detector did not collect data, as it was operated with HV turned off and the preamplifiers disabled to protect the FE chips from possible large charge depositions due to unstable beam configuration conditions.~After the LHC incident of September $\mathrm{19^{th}}$, taking data with cosmic muons became the first priority and the only source for physics data suitable for further detector calibration and alignment.
\subsection{Collecting First Cosmic Data}
The results presented in this paper are based on a data sample collected during combined ATLAS cosmic running in fall of 2008. During this running period, over 400,000 cosmic muon tracks passing through the pixel detector have been recorded. Cosmic data have been collected with the solenoidal magnetic field switched on and off. The collected dataset was used to finalize the commissioning of the pixel detector as well as to improve the detector alignment. The number of tracks with at least one pixel hit as a function of time, as well as the fraction of pixel modules included in data taking, is shown in Fig.\ref{fig:09_Running}.\begin{figure}[!htbp]
\centering
\includegraphics[width=80mm, clip=true]{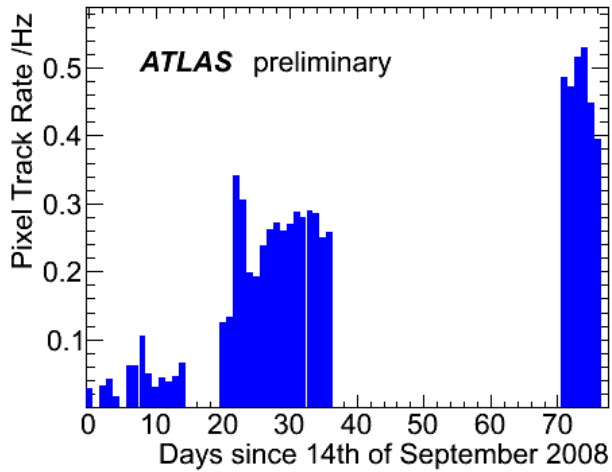}
\caption{The rate of tracks with at least one hit in the pixel detector as the function of time over the course of the cosmic data taking in fall 2008.} 
\label{fig:10_TrackRate}
\end{figure}

The Level-1 triggers for the pixel detector were provided by the {\it Resistive Plate Chambers} ({\it RPC}s) of the ATLAS muon spectrometer. Once a trigger was being received from the CTP, a wide readout window of $\mathrm{200~ns}$ (or 8 BC) was used by the pixel modules to ensure the readout of the entire detector, even if some parts of it were not timed correctly. In order to increase the dynamic range of the signal, the maximum front-end readout latency of 255 BC was used, which required delaying of the trigger signals from the CTP. Before the leaky cooling loops were disabled, about 96\% of the detector modules were being operated in the data-taking mode. The commissioning of the Level-2 trigger allowed for the track rate through the pixel volume to reach 0.3~Hz~\cite{bib:05_iskander_proceedings}.~Figure~\ref{fig:10_TrackRate} shows the rate of tracks with at least one pixel hit as a function of day for the running between September $\mathrm{14^{th}}$ and October $\mathrm{26^{th}}$, 2008.~The rate was increased substantially in early October, mostly due to significant improvements in the RPC trigger timing and the use of the track triggers at the Level-2. In November, a new L1 trigger of the TRT was developed, improving the track rate further to about $\mathrm{0.5~Hz}$, which is consistent with the anticipated rate of tracks crossing the detector.
\subsection{Preliminary Alignment}\begin{figure}[!hp]
\centering
\includegraphics[width=80mm, clip=true]{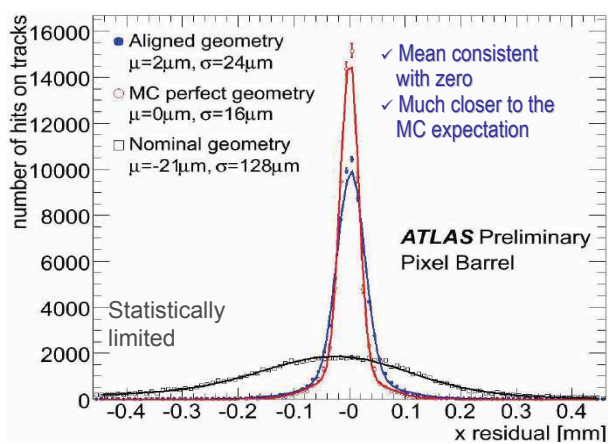}
\caption{Comparison of the residuals in $x$ precision direction, comparing the aligned detector geometry to the unaligned production geometry. While the result is statistically limited, the aligned geometry approaches the ideal, simulated alignment.} 
\label{fig:11_Resolution}
\end{figure}Alignment is a crucial task for ensuring the correct interpretation of detector readout, which is important for successful data collection.~The general approach to the alignment procedure on the most basic level is concerned with the minimization of the residuals, which are the differences between the real detector geometry with respect to the ideal designed geometry. These differences are caused by slight production imperfections and assembly tolerances, and have to be properly accounted for.

One of the most characteristic plots pertaining to pixel detector alignment (Fig.\ref{fig:11_Resolution}) illustrates one of the latest results of the alignment applied to the barrel of the pixel detector, showing that the obtained aligned geometry is getting very close to the simulated idealistic prediction with respect to the uncorrected physical detector, while the mean of the residual misalignment is consistent with zero. The obtained spatial resolution result of $\mathrm{24~\mu m}$ in the $r-\phi$ precision direction is statistically limited by the present size of the cosmic data sample.

Detailed studies of the pixel barrel geometry are ongoing, in the course of which a significant bowing effect with a {\it sagitta} parameter of up to $\sim \mathrm{400~\mu m}$ has been found in a few staves, and has been correctly accounted for in the detector offline profile~\cite{bib:05_iskander_proceedings}.

Currently, there are several algorithms in place to provide calculation and correction for the pixel detector mis-alignment.~The particulars of various alignment procedures deserve much more scrupulous attention, and are reviewed in more detail elsewhere~\cite{bib:09_alignment}. 
\subsection{Highlights of the Physics Performance}
\begin{figure}[h]
\centering
\includegraphics[width=80mm]{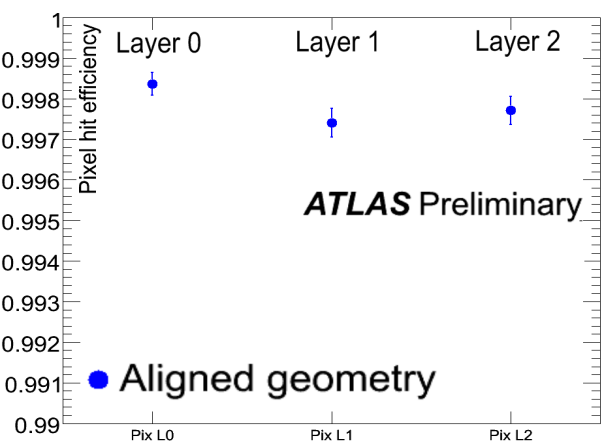}
\caption{Hit efficiency for 98\% of all pixel modules in barrel layers of the pixel detector using cosmic data collected in fall 2008.} 
\label{fig:12_Track_Efficiency}
\end{figure}
The cosmic ray data sample is being extensively analyzed to produce additional calibration measurements. In particular, the alignment improvements reflect on the overall physics performance of pixel detector. The efficiency of attaching hits on tracks in all active modules in the pixel barrels is shown in Fig.\ref{fig:12_Track_Efficiency}. It is measured to be about 99.8\% in all three layers using the official alignment algorithm.~The disabled modules are not being included in the denominator of the efficiency calculation ratio.
\begin{figure}[h]
\centering
\includegraphics[width=80mm]{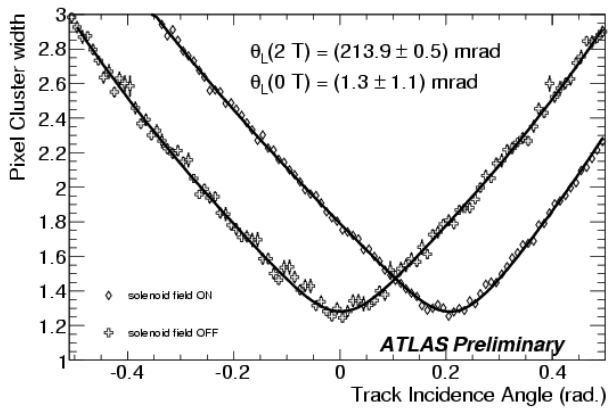}
\caption{The width of the pixel cluster versus the track incidence angle in azimuthal direction for pixel clusters on tracks. The measurement results with both the solenoidal magnetic field on and off are presented.} 
\label{fig:13_Lorentz_Angle}
\end{figure}
Another measurement essential to performance validation is the measurement of the Lorentz angle. The Lorentz effect produces a systematic shift between the position of the signal induced on the electrodes and the position of the track. While this shift is practically absorbed by the alignment correction, the knowledge of the Lorentz angle will help understanding the alignment corrections and their time dependence. In addition, the Lorentz effect is expected to change the angular dependence of the spatial resolution. Figure~\ref{fig:13_Lorentz_Angle} presents the plot of the pixel cluster width versus the incidence angle of the track, corresponding to the two different values of the solenoidal magnetic field. The minima of these distributions occur at the Lorentz angle, which depicts the deflection of the charge carriers in the magnetic field. This angle is consistent with zero when the magnetic field is off, and is about $\mathrm{214~mrad}$ for the full magnetic field strength of $\mathrm{2~T}$.~This result is consistent with the simulated expectation within 5\%~\cite{bib:10_lorentz_angle}.
\section{Summary}
After an intensive commissioning effort and cosmic-ray data taking period over the year 2008, (more than 96\% of) the ATLAS Pixel detector is now tuned and calibrated. The cosmic-ray data sample is being used for further improvement in calibration, fine-tuning of the most recent simulations, and ongoing studies of the detector performance. After the successful upgrade and additional commissioning work done on the cooling system, the entire detector has been re-calibrated. Starting June $\mathrm{19^{th}}$, 2009, more cosmic-ray data are being collected.
 
With a pixel hit efficiency over 99.7\% in the enabled modules, noise occupancy at the level of $10^{-10}$, and reaching towards a spatial hit resolution on the order of $\mathrm{20~\mu m}$ in the $r-\phi$ precision direction in the barrel layers, the ATLAS silicon pixel detector is performing well up to the expectations. The pixel detector is ready to collect physics data produced by beam collisions in the LHC, to be delivered by the end of 2009. 
\bigskip
\begin{acknowledgments}
The presented material and this article are the result of numerous individual contributions made by many members of the ATLAS Pixel Detector collaboration and the ATLAS Inner Detector group. I would like to kindly thank all of them.
\end{acknowledgments}

\bigskip 

\end{document}